# DoS Attacks at Cooperative MAC

Authors: [1]Zeeshan Haider, Kiramat Ullah and Tauseef Jamal

[1]ARID University Rawalpindi, Pakistan.

luckier19@gmail.com

Abstract—Cooperative networking brings performance improvement to most of the issues in wireless networks, such as fading or delay due to slow stations. However, due to cooperation when data is relayed via other nodes, there network is more prone to attacks. Since, channel access is very important for cooperation, most of the attacks happens at MAC.

One of the most critical attack is denial of service, which is reason of cooperation failure. Therefore, the cooperative network as well as simple wireless LAN must be defensive against DOS attacks.

In this article we analyzed all possible of DoS attacks that can happen at MAC layer of WLAN. The cooperative protocols must consider defense against these attacks. This article also provided survey of available solutions to these attacks. At the end it described its damages and cost as well as how to handle these attacks while devising cooperative MAC.

# 1. Introduction

In Cooperative Networks the data is relayed via helper nodes. Instead of sending data directly to destination, it is sent to relay where the relay can forward it on behalf to source to the destination. Let's consider if the relay is non processive or move away in meanwhile. The data is lost simply. Therefore, the cost of cooperation is DoS attack. If we prevent our MAC from DoS, we can get the benefits of cooperation.

DoS are such type of attack on a network that's aim is to exhaustedly usage of the network or system resources, which hold down forcedly the usage of network or system resources for the legitimate users. As shown in fig 1.

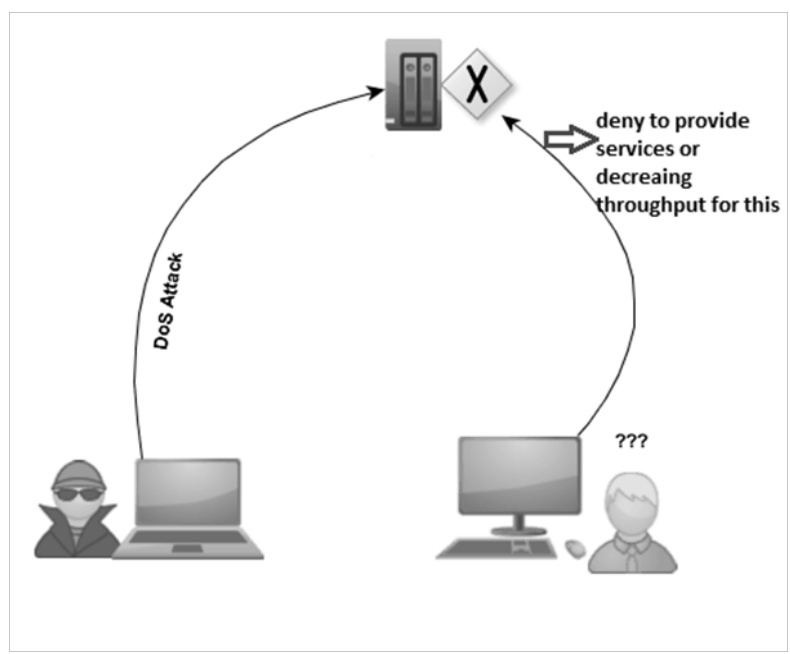

**Figure 1 DoS Attack**

A Denial-of-Service attack DoS is such kind of attack which targeting the accessibility of network system resources for other legitimate users [1]. In other kinds of attacks, the information is stolen or changes the data but DoS attack aim is slow down or takes down system resources for other users. The attackers' goals are diverse; he does that for simple fun or financial gain and ideology. The first step in denial of service (DoS) attack is generating high rate malicious traffic [2]; direct that malicious traffic flow towards victim network or resources' and consuming computing resources of target exhaustively. Therefore legitimate users are not able to access the system resources [3].

DoS attacks influence all organizations of the world. They can target all 7 layers of OSI model from physical layer a to the application layer. The difficult part of DoS attack is detection because traffic type seems legitimate traffic to the system resources [4].

There are two types of DoS attacks. A (non-distributed) DoS attack and distributed DoS attack. In non-distributed DoS attack, an attacker uses a single machine's to overwhelm another machine. If target machine powerful then this type of attack doesn't affect target system. While in distributed DoS case, the attacker originates from multiple computers simultaneously, focus on single or multiple machines, therefore, causing the victim's resources exhaustion [5].

## 2. How does an attack work?

1) The first attacker chose to find the goals and system for the attack. Then he discovers the target network and calculated all the limitations of network and system resources.
2) After first phase an attacker floods company's network or system with useless and malicious information [6].
3) Since Network and system can only handle a limited amount of traffic and an attacker overloads the targeted system with the unlimited amount of traffic.
4) Denial-of-service attacks disable the computer or the network partially or completely depending on the nature of the enterprise [7].

For example in authentication flood, the users send an authentication request to AP, AP respond with approval if there is space for approving. If the user has malicious intention then he can flood the AP by sending the flood of authentication request which causes AP to respond and hence others nodes of the network face DoS [8].

## 3. Attack Types

1. Packet Internet or Inter-Network Groper (Ping) Flood Attack or (ICMP echo)
2. (synchronization)SYN Flood Attack (DoS attack)
3. DDoS Attack (Distributed SYN Flood)
4. Land Attack (Local Area Network Denial)
5. Authentication request flood
6. Association request flood

7. CTS Flood attack
8. RTS DoS Attack
9. Beacon Flood

## *3.1 Ping Flood Attack (ICMP echo)*

In Ping flood attack, the attacker focus is network bandwidth. An attempt by an attacker on a network focus is bandwidth, fill a network with ICMP echo request packets in order to slow or stop legitimate traffic going through the network. As shown in fig 2.

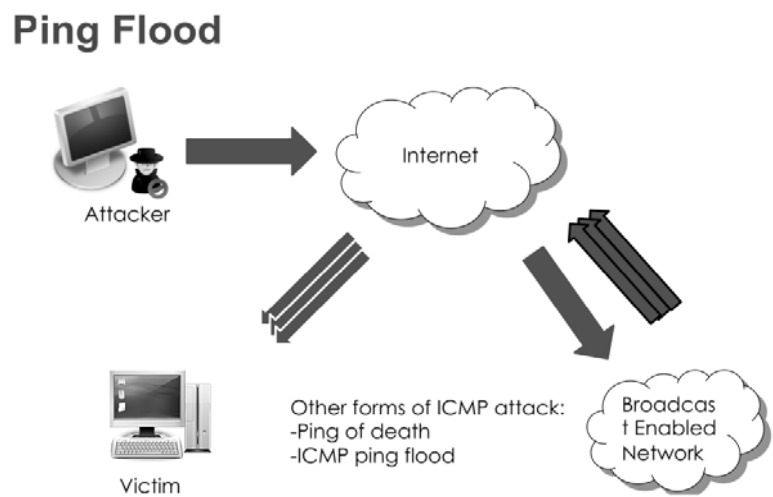

Figure 2 Ping Flood Attack

Ping is a basic network program, which used for checking that system is alive to receive data or not. When a system receives the Ping message, the system must reply if it alive and active. Ping flood is also known as ICMP flood, To create DoS in the network, the attacker sends thousands of ping messages to victim node and victim node just only busy with responding that he is alive. At that time victim system are not able to process the other nodes information. Victim system is even not able to receive other data in worst case scenario. [10]

## *3.2 SYN Flood Attack*

SYN messages are exchanges when a client needs to connect to a server in TCP. The user sends an SYN message, in response server send back SYN-ACK message [11]. In SYN flood attacker sends so many SYN requests that the system is notable for other nodes to respond. Since the server is busy with the reply to malicious SYN message and legitimate users are in the waiting stage. As explained in fig 3. [9]

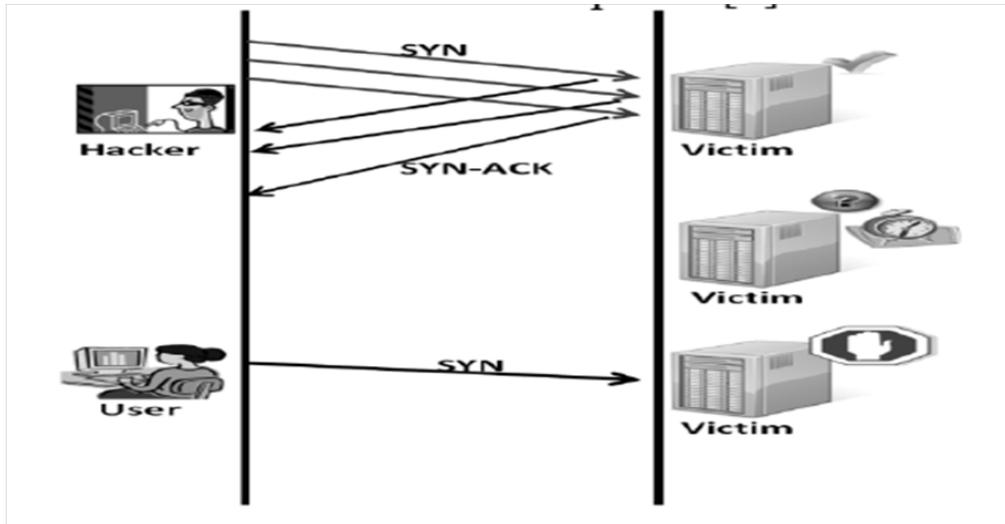

Figure 3 SYN Flood Attack

## 3.3 DDoS Attack

Distributed Denial of Services (DDoS) is such kind of DOS attack there are many step stone systems are used for generating malicious traffic and after that directed the flow of malicious traffic to the victim system and that cause a Denial of Service (DoS) attack. As shown in fig 4

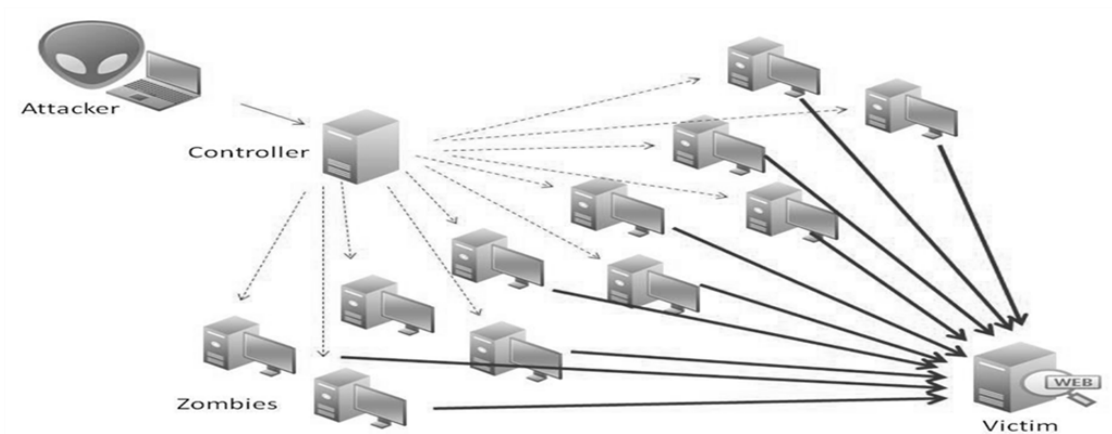

Figure 4 DDoS Attack Flow of traffic

### 3.3.1 How DDoS Attacks Work

There are three steps to launch the DDoS attack [12]. The main goal of the attacker is launching a large traffic and makes that flow direction towards victim system. For that, he first compromised many other systems called zombies. They are compromised using Trojans, infected system with malicious software and getting control of that zombie system. Using zombies having many advantages for the attacker, it's become impossible to block all zombies IPs addresses after detection. Each zombie generated traffic and direct that flow towards the victim. Even zombies detected attacker ID can't be detected. [13]

To handle zombies there is a controller in the second step. This may be also a compromised system or a system used by attacker temporarily. Controller, take instruction from an attacker, like how many zombies would be involved and for how much time, also malicious traffic format. Even victim find the controller, attackers ID are still hidden from the victim. The zombies and controller are used as step stone in the above two phases. The third step is traffic directed towards the victim [14].

### 3.3.2 Types of DDoS Attacks

There are many types of DDoS attacks. Common attacks include the following:

- **Traffic attacks:** In traffic attacks, the DDoS traffic is legitimate traffic like TCP, UDP, and ICMP. It's impossible for the victim to distinguish among malicious traffic and legitimate traffic because traffic pattern is same as like legitimate traffic. That's preventing legitimate user to access the system or network [15].

- **Bandwidth attacks:** In that kind of attack attacker's aim is bandwidth only. So he fills the bandwidth with junk data. Traffic can be easily distinguished by victims but the amount of traffic is so much that it can't be handling [16].

- **Application attacks:** In application attack, the attacker exploited the application layer and resource unavailable for legitimate users after malicious traffic. Application layers distributed data to system resources.

### *3.3.3 Land Attack (Local Area Network Denial)*

- It's an old kind of attack. In land attack, the attackers send malicious packets such that it has the same source and destination address. Both host and source addresses are victim addresses. It's mostly used in local area networks. The victim system is lock up after getting that packets and response to itself and loop continue until system detected or shutdown. As shown in fig 5.

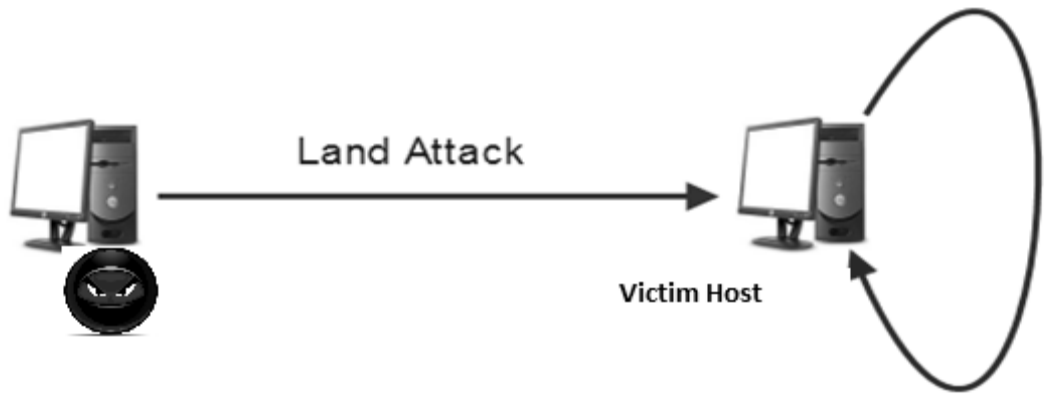

**Figure 5 Land Attack (Local Area Network Denial)**

### *3.3.4 Authentication request flood*

- A node after listening bacon sends authentication request to AP, to associate itself with AP.

- AP maintains a state table, where there is the list of authenticated nodes.

- There are two kinds of effects of such DoS attack, First AP affected, because commit its normal operation and serve the request, when the request is too much, AP only will do the job maintaining the state table. The second effects are legitimate users when state table is filled by malicious requests, there would be no space for accepting more legitimate requests. State table also has limitations. Shown in fig 6.

- In that kind of attack attacker first, need to spoof the MAC of others node. So it's little difficult to launch if there is the proper mechanism of protection for MAC addresses. [17]

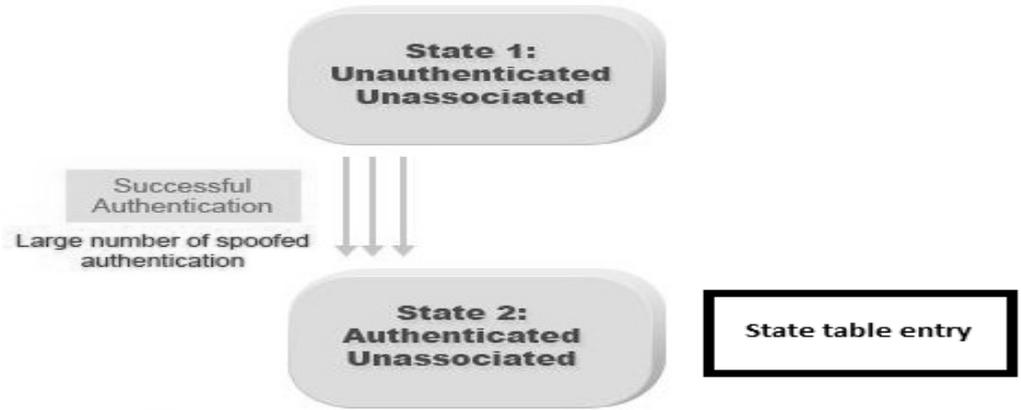

Figure 6 Authentication request flood

### 3.3.5 Association request flood

- After authentication, there is association step, in association step AP associate a client and make the entry in the association table. But this association is also vulnerable to DoS. There is de-authentication packet after authentication from AP if that de-authentication packet is spoofed and an attacker crack passwords then he can also reach to the association table. As shown in fig 7.

- That table also has limits and if requests are beyond the limit of an associated table, there would defiantly a DoS attack.

- It's harder to launch, because of the authentication step. An attacker must cross the authentication step [18].

- 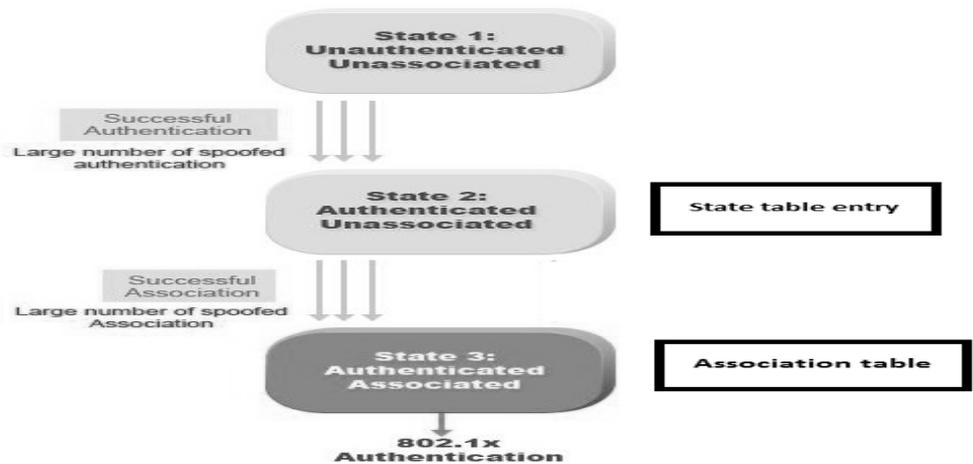

Figure 7 Association request flood

-

## 3.3.6 CTS Flood attack

- IEEE 802.11 set standard for wireless networks. As we discussed in the previous chapter, first there is RTS, followed by CTS, then DATA and ACK frame.

- Other nodes after listening CTS just update NAV and stay in quite a mood and start sensing media after CTS maintained time duration.

- This behavior can be exploited by an attacker, if an attacker sends CTS to others after the interval to others node, other nodes would be in quite a state after receiving.

- If the sending malicious CTS are back to back, no other node is able to send data. As shown in fig 8.

- There is also possible that CTS sender node increase the duration and nodes goes in the quiet state for the extra time.[17]

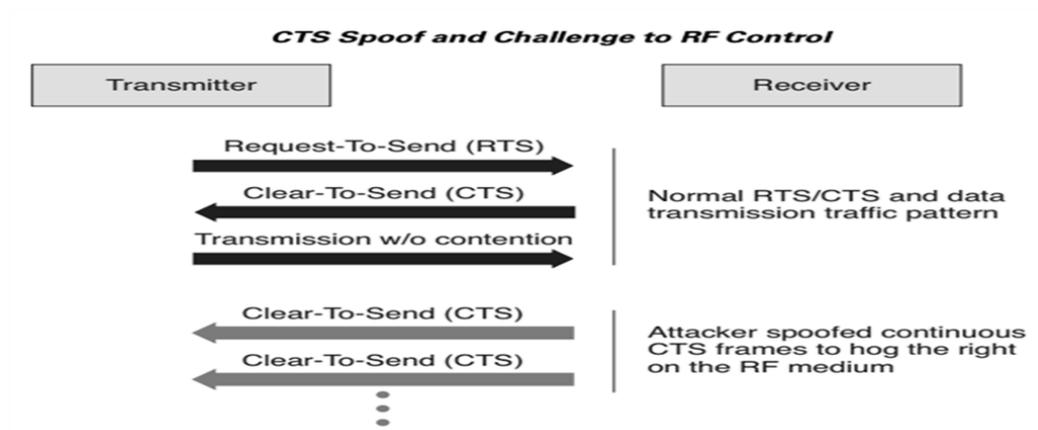

**Figure 8CTS Flood attack**

## 3.3.7 RTS DoS Attack

- RTS frame includes Frame Control, Duration, RA, TA, and FCS. By sending RTS frames mentioning large transmission duration, an attacker reserves the wireless medium for the overdue time and forces others wireless stations sharing the RF medium to delay their transmissions. As shown in fig 9.[18]

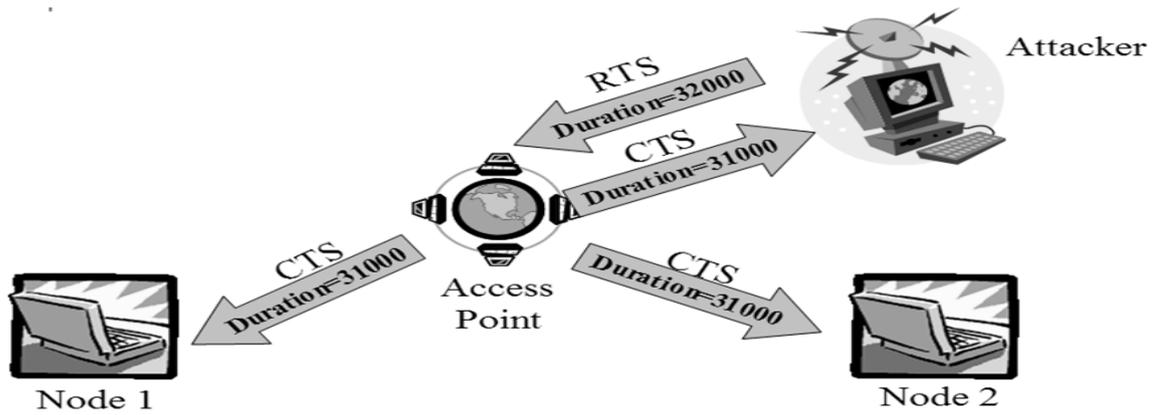

Figure 9 RTS Flood

### 3.3.8 Beacon Flood

Wireless clients can detect the presence of access points by listening for the beacon frames transmitted from APs. Beacon flood is launched by an attacker in such way, that first he generates thousands of malicious beacons around legitimate [20] AP that made difficult for the individual station to find the legitimate AP for the association. As shown in fig 10.

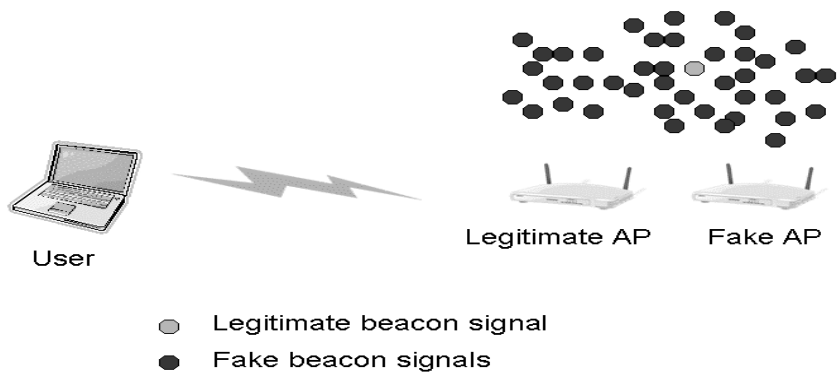

Figure 10 Beacon Flood

## 4. Damage & Costs

1. **Other affecting:** There are many costs associated with denial-of-service attacks. Like an attacker target the server, when server down, it does not only effect the server but also other users and sites associated with that victim server [19].

2. **Bandwidth wastage:** Network resources are shared among many stations. Like bandwidth. If attacker launches DDoS attack it does not only affect the target because of wastage of bandwidth and that also slow down the activity of non-victim systems [21].

3. **Extra network channels:** To detect the attack users must use extra resources only to handle and prevent their system from such kind of attacks. Like emailing, making logs etc.

4. **Insurance & Bandwidth cost:** As in international market we pay per byte. In DoS attack case the traffic is very high from normal traffic and that also increases the bandwidth cost.

## 5. How to handle DoS

- **Protecting:** The first step should be protected in such kind of attack, protection mechanism should be installed by ISP, and there should be an agreement between ISP, an insurance policy. Most of the people do that after learning a lesson.

- **Detecting:** If you detect properly then you would be able to respond accurately. For detection, there should be proper check and balance on log system, traffic pattern, updated blacklist and all updated detection software. The attacker use different mechanism to launch the attack. So maybe detection not helps out in some kind of attacks [22].

- **Reacting:** Reaction step comes when there is no proper protection and detection mechanism. In that step there would some technical steps which are mostly implemented, are informing ISP, start backup system and moving data to the backup system, decreasing the incoming traffic, applying available data content filters on incoming traffic, redirecting traffic, shut downing after data is moved. [23]

## 6. Available Solutions

- The DoS attacks at the MAC layer discussed here are very common in the IEEE 802.11 standard networks.

- The attacker exploited mostly the non-implementation of the authentication method for management and control frames.

- Mostly available solutions are cryptographically protecting of management and control frames. In that method first step is finding the vulnerability on the basis of cryptography and then the possible solution to mitigate these attacks.

- IEEE made an amendment to the original standard IEEE 802.11 and releases a new standard 802.11w.It included the security features for management frames like data confidentiality, data origin authenticity, and replay protection [27].

- But for control frames, there are still no cryptographic protection schemes at the MAC layer. So control frames are still vulnerable to DoS attack. An attacker can easily exploit the control frame by spoofing them and then use for resource exhaustion.

- The de-authentication vulnerability, in particular, can be fixed by authenticating control frames explicitly [26].

- De-authentication flooding, in particular, can be mitigated by delaying the effect of requests.

- In RTS DoS attack, the network performance can be restored back by Reevaluate RTS Duration (RRD) technique [25].

- MAC address spoofing can be protected if there is incrimination mechanism implanted in firmware in each node. When a node sends its MAC address there would incrimination after next frame by sender node. Since firmware functionality of wireless card can't be changed by an attacker. The receiver will only accept and response such frames which have incremented MAC [24].